\documentclass[]{article}
\usepackage{xcolor}
\usepackage{amsmath}
\usepackage{amsfonts}
\usepackage{amssymb}
\usepackage{mathtools}
\usepackage{todonotes}
\usepackage[skip=2pt]{caption}
\usepackage{authblk}
\usepackage{float}
\usepackage{amsthm}
\usepackage{relsize}

\pdfoutput=1 

\newcommand\blfootnote[1]{%
	\begingroup
	\renewcommand\thefootnote{}\footnote{#1}%
	\addtocounter{footnote}{-1}%
	\endgroup
}

\theoremstyle{remark}

\title{Mathematical Foundations of Regression Methods
	for Approximating the Forward Initial Margin}

\author[1]{Lucia Cipolina-Kun}
\author[2]{Simone Caenazzo}
\author[2]{Ksenia Ponomareva}
\affil[1]{University of Bristol. Department of Electrical and Electronic Engineering}
\affil[2]{Riskcare Ltd., London}

\begin{document}

\maketitle

\begin{abstract}
	
\blfootnote{corresponding author: lucia.kun@bristol.ac.uk}

The modelling of forward initial margin poses a challenging problem as it requires the implementation of a nested Monte Carlo simulation, which is computationally intractable. Abundant literature has been published on approximation methods aiming to reduce the dimensionality of the problem, the most popular ones being the family of regression methods. This paper describes the mathematical foundations, on which these regression approximation methods lie. Mathematical rigor is introduced to show that, in essence, all methods are performing orthogonal projections on Hilbert spaces, while simply choosing a different functional form to numerically estimate the conditional expectation. The most popular methods in the literature so far are covered here. These are polynomial approximations, Kernel regressions and Neural Networks.

\end{abstract}

\section{Introduction}

Initial margin (IM) has become a topic of high relevance for the financial industry in recent years. The relevance stems from its wide implications to the way business will be conducted in the financial industry and in addition, for the need of approximation methods to calculate it. We refer to initial margin as a collateral posted and/or received in the context of OTC transactions. Its main purpose is to reduce future exposure in the event of a counterparty default. It differs from \emph{variation} margin as:

\begin{itemize}
    \item It is meant to cover losses over a certain time horizon, called Margin Period of Risk (MPOR) and often assumed to be 1 week or more;
    \item It is a \emph{segregated} amount, that must be held into a third-party account (such that of a custodian bank) and cannot be pledged as offset to any other liability.
\end{itemize}

\medskip
\noindent The initial margin amount is calculated at the netting set level following a set of rules depending on whether the trade has been done bilaterally \cite{ISDA-SIMM} or facing a Clearing House \cite{Henrard}. The calculation is performed on a daily basis and carried on throughout the life of the trade. Since each netting set consumes initial margin throughout its entire life, financial institutions need resources to fund this amount in the future, given that the initial margin received from its counterparty cannot be rehypothecated. To estimate the total need for funds, each counterparty needs to perform a forecast of the initial margin amount up to the time to maturity of the netting set, this forecast of initial margin consumption is denoted as \textit{forward initial margin}.

\medskip
\noindent Forecasting initial margin amounts is a computationally challenging problem, as an exact calculation requires a full implementation of a nested Monte Carlo simulation which is computationally onerous in most practical settings.  Practitioners and academics have proposed several approximation methods to reduce the dimensionality of the problem, at the cost of losing a tolerable degree of accuracy. Some examples are the Chebyshev polynomials proposed by \cite{Ignacio}, automatic differentiation methods proposed by \cite{Fries}, Gaussian processes \cite{Crepey}, and regression methods \cite{Justin}. Among all, regression methods have been the most widely spread due to their simplicity, re-usability of the Bank's legacy Monte Carlo engines, and reasonable results. Inspired by early works of Longstaff-Schwartz \cite{LS}, several authors have proposed some version of the polynomial regression,  \cite{Anfuso, Caspers, Justin}, or the Kernel regressions proposed by \cite{Andersen-Exposure, Justin, Dahlgren}. Most recently, estimation by Neural Networks has been proposed by \cite{Hernandez} and \cite{DeepLearning}. 

\medskip
\noindent The financial literature on forward initial margin has focused on the implementation aspects of regression methods from a practical point of view. The current literature is mainly focused on the the applicability and performance of the different methodologies, without dwelling on mathematical formalities. It is here where our paper finds its value, i.e. by filling the current gap between theory and practice. 

\medskip
\noindent The contribution of this paper can be summarized as follows:

\begin{itemize}
    \item To the best of our knowledge, the present work is the first one that aims at describing the rigorous mathematical framework behind regression methodologies for determining the forward initial margin. 
    
    \item Unlike the current literature, our emphasis is placed on the mathematical conditions needed for regression methods to be accurate unbiased estimators of the forward initial margin. In particular, we emphasize the Markovian property and the necessary conditions placed on the Mark-to-market processes. Note that MtM processes are not, in general, guaranteed to have such properties.
    
    \item The paper draws a clear demarcation line between assumptions, necessary conditions and empirical calls, paying particular attention to the choice of regressor variables. 
\end{itemize}

The underlying mathematical framework required to analyze the forward initial margin computation problem from a theoretical standpoint is presented here. 

\medskip
\noindent The rest of this paper is organized as follows. Section  2 translates the practitioner's definition of the forward initial margin into a formal probability setting. Sections 3-4 develop the assumptions needed to fit the forward initial margin problem into a classical problem of orthogonal projections into Hilbert spaces. The last section concludes by linking back the formal theory presented with the practitioner's suggested approaches. 

\section{Description of the Forward Initial Margin Problem}

\subsection{Definition of Forward Initial Margin}

\noindent A Bank engaged in OTC transactions needs to calculate the total cost associated with the posting of initial margin to a counterparty over the life of a netting set. Since this quantity evolves in time, it is necessary to consider a forward diffusion model, together with a probability space and the relevant discounting and funding rates into the equation. 

\medskip
\noindent Consider a filtered probability space $(\Omega, \mathcal{F},\left\{{\mathcal {F}}_{{t}}\right\}_{{t\geq 0}}, \mathbb{P})$ where $\Omega$ is the set of all possible sample paths $\omega$ 
, $\mathcal{F}$ is the sigma-algebra,$\left\{{\mathcal {F}}_{{t}}\right\}$ is the filtration of events at time $t$ and $\mathbb{P}$ the probability measure defined on the elements of $\mathcal{F}$. Following the definitions in \cite {Green}, the total cost of posting forward initial margin over the entire life of a netting set is referred to as the \textit{Margin Valuation Adjustment} as defined below.

\begin{equation}
\label{eq:MVA1}
MVA_C(t) = \mathbb{E}_{t}\left[ \int_{t}^{T} ((1-R_C)\lambda_B(u)-S_I(u)) e^{\int_{t}^{u}-(r(s)+\lambda_B(s)+\lambda_C(s))ds}\mathbb{E}_{u}\left[IM_C(u)\right] du \right]  
\end{equation}

\noindent where, $\mathbb{E}_{t}$ denotes an expectation conditional to information at time $t$, i.e. $\mathbb{E}_{t}\left[\dots\right] = \mathbb{E}\left[\dots \mid \mathcal{F}_t\right]$
, $R_C(t)$ is the recovery rate of the counterparty related to netting set $C$,  $T$ is the final maturity of netting set $C$, $S_I$ is the funding spread experienced by the bank when borrowing the initial margin amount, $r(t)$ is the risk-free rate (e.g. an OIS rate) at time $t$, $\lambda_B(t),\lambda_C(t)$ are the financing rates (i.e. spreads over risk-free rate) at time $t$ of the Bank and the counterparty related to netting set $C$, respectively, $IM_C(t)$ is the initial margin posted by the bank at time $t$ against netting set $C$.

\medskip
\noindent The intuition behind this formula is that it integrates three components:

\begin{itemize}
    \item The cost of funding forward Initial Margin for the bank;
    \item The forward Initial Margin amounts;
    \item The discounting of forward Initial Margin amounts, which also includes the impact of early default as characterised by the bank's and counterparty's funding spreads.  
\end{itemize}

\noindent Note that we have used the concept of netting set and counterparty. In general, a netting set is an ensemble of trades for which netting is allowed - therefore there is a one-to-many relationship between counterparty and netting sets, as one counterparty may hold more than one netting set when facing a bank. Furthermore, note that the expectation $\mathbb{E}_{t}$ is taken on the whole integral. This is because certain simulation setups, like the ones in \cite{AlbGimPie} and \cite{AlbaneseKVA}, explicitly model credit factors that lead to stochastic credit spreads and survival probabilities. If the simulation setup instead considers deterministic credit factors, the formula can be simplified as follows: 

\begin{eqnarray*}
MVA_C(t) &=&  \int_{t}^{T} ((1-R_C)\lambda_B(u)\\
&-&S_I(u))e^{\int_{t}^{u}-(\lambda_B(s)+\lambda_C(s))ds}\mathbb{E}_{t}\left[\mathbb{E}_{u}\left[IM_C(u)\right]e^{-\int_{t}^{u}r(s)ds}\right] du  
\end{eqnarray*}
 
\medskip
\noindent Here the quantity of interest is $\mathbb{E}_{t}\left[IM(t)\right]$, which is a short for $\mathbb{E}_{t}\left[IM_C(\omega,t, t+\delta_{IM})\right]$. This is the expectation of the initial margin at time $t$, taken across all random paths $\omega$, conditional on the realization of the risk factors at time $t$. The forward initial margin is a path-wise random variable defined as \footnote{This is the simplest definition of forward initial margin, for alternative definitions see \cite{RiskBook}.}

\begin{equation}
\label{eq:IM}
P(\Delta_t \leq IM(\omega,t,t+\delta_{IM}) \vert \mathcal{F}_t) = p
\end{equation}

\medskip
\noindent Where $p$ denotes the desired quantile (usually the $99\%$ quantile),  $\Delta(\omega,t,\delta_{IM})$ is defined (in its simplest form\footnote{This is the simplest definition of netting set PnL, for further including cash flow adjustments see \cite{RiskBook}.}) as the change, conditional on counterparty default, in the netting set's value (i.e. profit and loss (PnL)) over a time interval $(t,t+\delta_{IM}]$.
    

\medskip
\noindent The following sections describe the computational methods developed in the current literature to estimate this quantity.

\subsection{The Brute Force Approach to Calculate the Forward Initial Margin}

\noindent From a computational point of view, equation \ref{eq:MVA1} can be abstracted to a nested Monte Carlo problem, where two expectations are chained

\begin{equation*}
\label{eq:NestedMC}
\mathbb{E}_{t}\left[f\left(V(\omega,t),\mathbb{E}_{t} \left(IM(\omega,t, t+\delta_{IM}) \vert \mathcal{F}_t\right)\right) \right]
\end{equation*}

\noindent The most exact method to calculate the forward initial margin requires a \textit{brute force} implementation of a nested Monte Carlo simulation where for each sample of $V(\omega,t)$, one would draw i.i.d samples of $V(\omega,t+\delta_{IM})\vert V(\omega,t)$ . The left side of Figure \ref{fig:nested-mc1} exemplifies the situation. For a given point in time, $t_1$, an \textit{outer} Monte Carlo simulation for the values of $V(\omega,t_1)$ and an \textit{inner} Monte Carlo simulation of the forward Probability Density Function (PDF) of $\left(V(\omega,t_1+\delta_{IM})\mid \mathcal{F}_{t_1}\right)$ have been performed. The difference in the value of the netting set between $t_1$ and $t_1+\delta_{IM}$ time steps is $\Delta(\omega,t_1,\delta_{IM})$. Since the PDF distribution of $(V(\omega,t_1+\delta_{IM})\mid \mathcal{F}_{t_1})$ is known, it is possible to obtain the PDF distribution of $\Delta(\omega,t_1,\delta_{IM})$.  Then, one should take the $\mathcal{Q}_{99}$ to obtain the forward initial margin. Note that the number of scenarios on the tail should be sufficiently large to allow a reliable quantile estimation.

\medskip
\noindent Such a \textit{brute force} calculation of the forward initial margin is computationally onerous given that the number of operations increase exponentially. This type of complexity is generally not possible to afford in a real-time live trading scenario, thus requiring approximation methods.

\section{Summary}
The table below provides advantages and disadvantages from the practitioners' point of view of the methods covered.

\begin{center}
\begin{tabular}{|p{1.5cm}|p{3cm}|p{6cm}| }
\hline
 \textbf{Method} & \textbf{$m(X_t,\hat{\beta})$}&\textbf{Pros and Cons}  \\ 
 \hline
 Linear  Maps&\begin{tabular}{@{}p{3cm}@{}} \\ $\sum_{i=0}^N \hat{\beta_i}\cdot \phi_i(X_t)$\\  \\$\hat{\beta}\in R^N$ \\ \\$\phi_i$ -  linear basis functions\\ \\ N - degree of polynomial \end{tabular}& \begin{tabular}{@{}p{6cm}@{}} {\color{green}+} Same few calibrated parameters are used across all simulation scenarios and time steps.  \\
 {\color{green}+} Computationally cheap compared to the two other methods.\\
 {\color{red}-} Large oscillations in the calibrated parameters can be observed due to daily change in IM value at $t_0$.\\
 {\color{red}-} In a distributed grid setup a small simulation run is required to pre-compute $\hat{\beta_i}$ prior to the main MC simulation.  \end{tabular}\\
\hline  
Kernel   Regressions (NW) &\begin{tabular}{@{}p{3.1cm}@{}} \\ $\sum_{i=0}^N \hat{\beta_i}\cdot \phi_i(X_t)$\\ \\$\hat{\beta_i}=\frac{K_h(x_0,x_i)}{\sum_{j=1}^{N}K_h(x_0,x_j)}$ \\  \\$\phi_i= y_i$ \\  \\N - number of neighbourhood pairs ($x_i, y_i$) \\ \\ $K_h$ - Kernel function \end{tabular} & \begin{tabular}{@{}p{6cm}@{}}
{\color{green}+} Computationally cheap \\
 {\color{green}+} Better empirical fit to data than linear regression. \\
 {\color{green}+} Ensures functional smoothness. \\
 {\color{red}-} Assumes parametric functional form.\\
 {\color{red}-} It could be challenging to obtain sensitivities to model parameters and inputs.   
 \end{tabular}\\
\hline  
Feed-Forward Neural   Networks &\begin{tabular}{@{}p{3cm}@{}} \\ $\sum_{i=0}^N \hat{\beta_i}\cdot \phi_i(X_t)$\\  \\$\hat{\beta_i} = W^i_L$, weight applied to the output of the neuron i in the final hidden layer\\ \\ $\phi_i(X_t) = A_L^i$ - activation of the neuron i in the final hidden layer   \\ \\N - number of neurons in the final hidden layer (plus 1 if incorporating bias) \end{tabular}&  \begin{tabular}{@{}p{6cm}@{}} {\color{green}+} Once trained, predicted IM can be calculated extremely fast because it only involves small scale matrix multiplication and is done at the portfolio level.\\
 {\color{green}+} $\hat{\beta_i}$ only need to be updated every quarter compared to daily in linear maps method. \\
 {\color{red}-} Training requires portfolio sensitivities for each MC path and time-step per SIMM model requirements. \\
 {\color{red}-} Data generation for training is computationally expensive and needs to be done offline.  \end{tabular}\\
\hline  
\end{tabular}
\end{center}

\end{document}